\documentclass[12pt]{article}
\usepackage{enumerate}
\usepackage{amsfonts}
\usepackage{amsmath}
\usepackage{amssymb}
\usepackage{amsthm}
\usepackage{latexsym}
\usepackage{color}

\def\H{\mathcal{H}}

\def\P{\mathcal{P}}
\def\S{\mathfrak{S}}
\def\F{\mathfrak{F}}
\def\C{\mathfrak{C}}
\def\T{\mathfrak{T}}
\def\B{\mathfrak{B}}

\newcommand{\Ran}{\mathrm{Ran}}
\newcommand{\id}{\mathrm{Id}}
\newcommand{\Tr}{\mathrm{Tr}}

\newcommand{\shs}{\hspace{1pt}}
\newcounter{defin}  \newcounter{lemma}  \newcounter{theorem}
\newcounter{proposition} \newcounter{corol}  \newcounter{remark} \newcounter{example}
\newenvironment{lemma}{\par\refstepcounter{lemma}     \textbf{Lemma \thelemma.} }{\rm\par}
\newenvironment{theorem}{\par\refstepcounter{theorem}     \textbf{Theorem \thetheorem.}\ }{\rm\par}
\newenvironment{proposition}{\par\refstepcounter{proposition}     \textbf{Proposition \theproposition.}\ }{\rm\par}
\newenvironment{corollary}{\par\refstepcounter{corol}     \textbf{Corollary \thecorol.} }{\rm\par}

\newenvironment{remark}{\par\refstepcounter{remark}     \textbf{Remark \theremark.}}{\rm\par}

\begin{document}

\title{Strong convergence of quantum channels: continuity of the Stinespring dilation and discontinuity of the unitary dilation}

\author{M.E.~Shirokov \\
Steklov Mathematical Institute, Moscow, Russia}
\date{}
\maketitle
%\vspace{50pt}
\begin{abstract}
We show that a sequence $\{\Phi_n\}$ of quantum channels strongly converges to a quantum channel $\Phi_0$ if and only if there exist a common environment for all the channels and a corresponding sequence $\{V_n\}$ of Stinespring isometries strongly converging to a Stinespring isometry $V_0$ of the channel $\Phi_0$.

We also give a quantitative description of the above characterization of the strong convergence in terms of the appropriate metrics on the sets of quantum channels and Stinespring isometries. As a result, the uniform selective continuity of the complementary operation with respect to the strong convergence is established.

We show discontinuity of the unitary dilation by constructing a strongly converging sequence of channels which can not be represented as a reduction of a strongly converging sequence of unitary channels.

The Stinespring representation of strongly converging sequences of quantum channels allows to prove the lower semicontinuity of the entropic disturbance as a function of a pair (channel, input ensemble).  Some corollaries of this property are considered.
\end{abstract}

\tableofcontents

\section{Introduction}

The Stinespring theorem provides a characterization of quantum channels -- completely positive trace-preserving linear maps between Banach spaces of trace-class operators \cite{St}. It implies  that any quantum channel $\Phi$ from a quantum system $A$ to a quantum system $B$ can be represented as
\begin{equation}\label{S-r}
\Phi(\rho)=\Tr_E V_{\Phi}\rho V^*_{\Phi},
\end{equation}
where $V_{\Phi}$ is an isometrical embedding of the input Hilbert space $\H_A$ into the tensor product of the output Hilbert  space $\H_B$ and some
Hilbert  space $\H_E$ typically called environment \cite{H-SCI,Wilde}.

It is natural to explore  continuity of the representation (\ref{S-r}) with respect to appropriate metrics (topologies) $D$ and $D'$ on the sets of quantum channels and  corresponding Stinespring isometries. Since the map $\Phi\mapsto V_{\Phi}$ is multivalued, the question of its continuity should be formulated in the following form: is it possible to find for any $\varepsilon>0$ such  $\delta>0$ that for any channels $\Phi$ and $\Psi$ $\delta$-close w.r.t the metric $D$ there exist corresponding Stinespring isometries $V_{\Phi}$ and $V_{\Psi}$ $\varepsilon$-close w.r.t. the metric $D'$?  This question can be also formulated in terms of  sequences of channels $\{\Phi_n\}$ converging w.r.t. the metric $D$ and corresponding  sequences of \emph{selective} Stinespring isometries $\{V_{\Phi_n}\}$ converging w.r.t. the metric $D'$.

If $D$ and $D'$ are, respectively, the diamond-norm metric on the set of quantum channels and the operator-norm metric on the set of isometries then the above continuity question is completely solved by Kretschmann, Schlingemann and Werner in \cite{Kr&W+,Kr&W}. They have shown that
\begin{equation}\label{star}
\textstyle\frac{1}{2}\|\Phi-\Psi\|_{\diamond}\leq \inf\|V_{\Phi}-V_{\Psi}\|\leq\sqrt{\|\Phi-\Psi\|_{\diamond}}
\end{equation}
for any channels $\Phi$ and $\Psi$, where the infimum is over all the isometries $V_{\Phi}$ and $V_{\Psi}$ from common Stinespring representations of these channels.

The diamond-norm metric is widely used  as a  measure of distinguishability between finite dimensional quantum  channels \cite[Ch.9]{Wilde},\cite{Kit}. But the topology (convergence) generated by the diamond-norm metric on the set of infinite-dimensional quantum channels is generally \emph{too strong} for analysis of real variations of such  channels \cite{SCT,W-EBN}. In this case it is natural to use the substantially weaker \emph{topology of strong convergence} on the set of  quantum channels defined by the family of seminorms $\Phi\mapsto\|\Phi(\rho)\|_1$, $\rho\in\S(\H_A)$ \cite{AQC}. The strong convergence of a sequence $\{\Phi_n\}$ of channels to a channel $\Phi_0$  means that
\begin{equation}\label{star+}
\lim_{n\rightarrow\infty}\Phi_n(\rho)=\Phi_0(\rho)\,\textup{ for all }\rho\in\S(\H_A).
\end{equation}
%The equivalent definition of the strong convergence and analysis of its properties are presented in \cite{Wilde+}.

In this paper we present a modification of the  Kretschmann-Schlingemann-Werner theorem  for the strong convergence topology on the set of quantum channels and the strong operator topology on the set of Stinespring isometries. This modification is based on using  the \emph{energy-constrained Bures distance} between quantum channels introduced in \cite{CID} and the \emph{operator E-norm}  generating the strong operator topology on bounded subsets of $\B(\H)$ (introduced in Section 3).

The modified version of the  Kretschmann-Schlingemann-Werner theorem allows to obtain a
characterization of the strong convergence of quantum channels in terms of the Stinespring representation. It states, roughly speaking, that the strong convergence of a sequence $\{\Phi_n\}$ of quantum channels  is equivalent to the strong (operator) convergence of a corresponding  sequence $\{V_{\Phi_n}\}$ of \emph{selective} Stinespring isometries. We also obtain a
characterization of the strong convergence of quantum channels in terms of the Kraus representation and prove the uniform selective continuity of the complementary operation $\,\Phi\mapsto\widehat{\Phi}\,$ with respect to the strong convergence of quantum channels.

By using the Stinespring representation (\ref{S-r}) it is easy to show  that any quantum channel $\Phi$ can be represented
as a reduction of some unitary (reversible) evolution of a larger quantum system. In the case $A=B$ this means that
\begin{equation*}%\label{ud-0}
  \Phi(\rho)=\Tr_E\shs U_{\Phi} \rho\otimes\sigma_0 U_{\Phi}^*,
\end{equation*}
where $\sigma_0$ is a pure state in $\S(\H_E)$ and $U_{\Phi}$ is a unitary operator on $\H_{AE}$ \cite{H-SCI,Wilde,Kraus}.
We analyse the question of selective continuity of the multivalued map  $\Phi\mapsto U_{\Phi}$  w.r.t. given metrics (topologies) on the sets of quantum channels
and of unitary operators.

By using the  Kretschmann-Schlingemann-Werner result mentioned above we prove the selective continuity of the map  $\Phi\mapsto U_{\Phi}$  w.r.t. the diamond norm
metric on the set of quantum channels and the operator norm metric on the set of unitary operators (Proposition \ref{d-n-ud}). It means  that any sequence of quantum channels converging w.r.t. the diamond norm \emph{can be represented} as a reduction of a sequence of unitary channels converging w.r.t. the diamond norm.

In the case of  strong convergence topologies we show  discontinuity of the map  $\Phi\mapsto U_{\Phi}$ by constructing a  strongly converging sequence  $\{\Phi_n\}$ of channels with Choi rank $2$ which can not be represented  as a reduction of a strongly converging sequence of unitary channels (Corollary \ref{ud-cr-c}). This discontinuity
means that some strongly converging sequences of channels \emph{have no sense} within the standard interpretation
of a channel as a reduced unitary evolution of a larger system.

The obtained characterization of the strong convergence in terms of the Stinespring representation can be applied to continuity analysis of different entropic and information characteristics of quantum channels. In the last section of the paper we use it to prove the lower semicontinuity of the entropic disturbance as a function of a pair (channel, input ensemble) w.r.t. the strong convergence of channels and the weak convergence of ensembles. This result is  derived from the lower semicontinuity of the entropic disturbance as a function of an input ensemble established in \cite{HQL}.

The lower semicontinuity of the entropic disturbance is used to prove the
closedness of the set of all quantum channels reversible with respect to a given family of input states. It also implies  continuity properties of
the output Holevo quantity of a quantum channel and of a complementary channel which can be treated as stability of this quantities  with respect to all physical perturbations of a channel.

\section{Preliminaries}

Let $\mathcal{H}$ be a separable  Hilbert space,
$\mathfrak{B}(\mathcal{H})$ the algebra of all bounded operators  on $\mathcal{H}$ with the operator norm $\|\cdot\|$ and $\mathfrak{T}( \mathcal{H})$ the
Banach space of all trace-class
operators on $\mathcal{H}$ with the trace norm $\|\!\cdot\!\|_1$. Let
$\mathfrak{S}(\mathcal{H})$ be  the set of quantum states (positive operators
in $\mathfrak{T}(\mathcal{H})$ with unit trace) \cite{H-SCI,Wilde}.

Denote by $I_{\H}$ the unit operator on a Hilbert space
$\mathcal{H}$ and by $\id_{\mathcal{\H}}$ the identity
transformation of the Banach space $\mathfrak{T}(\mathcal{H})$.

The \emph{Bures distance} between quantum states $\rho$ and $\sigma$ is defined as
\begin{equation}\label{B-d-s}
  \beta(\rho,\sigma)=\sqrt{2\left(1-\sqrt{F(\rho,\sigma)}\right)},
\end{equation}
where $F(\rho,\sigma)=\|\sqrt{\rho}\sqrt{\sigma}\|^2_1$ is the \emph{fidelity} of $\rho$ and $\sigma$ \cite{H-SCI,Wilde}. The following relations  between the Bures distance and the  trace-norm distance hold
\begin{equation}\label{B-d-s-r}
\textstyle\frac{1}{2}\|\rho-\sigma\|_1\leq\beta(\rho,\sigma)\leq\sqrt{\|\rho-\sigma\|_1}.
\end{equation}

The \emph{von Neumann entropy} $H(\rho)=\mathrm{Tr}\eta(\rho)$ of a
state $\rho\in\S(\mathcal{H})$, where $\eta(x)=-x\log x$ if $x>0$ and $\eta(0)=0$,
is a concave nonnegative lower semicontinuous concave function on the set $\S(\mathcal{H})$ \cite{H-SCI,W}.

The \emph{quantum relative entropy} for two states $\rho$ and
$\sigma$ in $\S(\mathcal{H})$ is defined as
$$
H(\rho\shs\|\shs\sigma)=\sum_i\langle
i|\,\rho\log\rho-\rho\log\sigma\,|i\rangle,
$$
where $\{|i\rangle\}$ is the orthonormal basis of
eigenvectors of the state $\rho$ and it is assumed that
$H(\rho\shs\|\shs\sigma)=+\infty$ if the support of $\rho\shs$ is not
contained in the support of $\sigma$ \cite{H-SCI,W}.\footnote{The support of a positive operator is the orthogonal complement to its kernel.}

If quantum systems $A$ and $B$ are described by Hilbert spaces  $\mathcal{H}_A$ and $\mathcal{H}_B$ then the bipartite system $AB$ is described by the tensor product of these spaces, i.e. $\mathcal{H}_{AB}\doteq\mathcal{H}_A\otimes\mathcal{H}_B$. A state in $\mathfrak{S}(\mathcal{H}_{AB})$ is denoted by $\rho_{AB}$, its marginal states $\mathrm{Tr}_B\rho_{AB}$ and $\mathrm{Tr}_A\rho_{AB}$ are denoted, respectively, by $\rho_{A}$ and $\rho_{B}$ (here and in what follows $\mathrm{Tr}_X$ denotes the partial trace $\mathrm{Tr}_{\mathcal{H}_X}$ over the space $\mathcal{H}_X$).\smallskip

A \emph{quantum channel} $\,\Phi$ from a system $A$ to a system
$B$ is a completely positive trace preserving linear map from
$\mathfrak{T}(\mathcal{H}_A)$ into $\mathfrak{T}(\mathcal{H}_B)$ \cite{H-SCI,Wilde}. For any  quantum channel $\,\Phi:A\rightarrow B\,$ the Stinespring theorem implies existence of a Hilbert space
$\mathcal{H}_E$ and of an isometry
$V_{\Phi}:\mathcal{H}_A\rightarrow\mathcal{H}_B\otimes\mathcal{H}_E$ such
that
\begin{equation}\label{St-rep}
\Phi(\rho)=\mathrm{Tr}_{E}V_{\Phi}\rho V_{\Phi}^{*},\quad
\rho\in\mathfrak{T}(\mathcal{H}_A).
\end{equation}
The space $\H_E$ is called \emph{environment}, its minimal dimension is called \emph{Choi rank} of the channel $\Phi$ \cite{H-SCI,Wilde}.\smallskip

In finite dimensions (i.e. when $\dim\H_A$ and $\dim\H_B$ are finite) the distance  between quantum channels from $A$ to $B$ generated by the diamond norm
\begin{equation}\label{d-norm}
\|\Phi\|_{\diamond}\doteq \sup_{\rho\in\S(\H_{AR})}\|\Phi\otimes \id_R(\rho)\|_1
\end{equation}
of a Hermitian-preserving superoperator $\Phi:\T(\H_A)\rightarrow\T(\H_B)$, where $R$ is any system, is widely  used  as a  measure of distinguishability between these channels \cite{Wilde,Kit,Paul}. It is topologically equivalent to the Bures distance
\begin{equation}\label{b-dist+}
\beta(\Phi,\Psi)=\sup_{\rho\in\S(\H_{AR})} \beta(\Phi\otimes \id_R(\rho),\Psi\otimes \id_R(\rho))
\end{equation}
between quantum channels $\Phi$ and $\Psi$, where $\beta(\cdot,\cdot)$ in the r.h.s. is the Bures distance between quantum states defined in (\ref{B-d-s}) and $R$ is any system. This metric is related to the notion of \emph{operational fidelity} for quantum channels introduced in \cite{B&Co}. It is studied in detail in \cite{Kr&W+,Kr&W}. In particular, it is shown in \cite{Kr&W} that the Bures distance (\ref{b-dist+}) can be also defined as
\begin{equation}\label{b-dist}
\beta(\Phi,\Psi)=\inf\|V_{\Phi}-V_{\Psi}\|,
\end{equation}
where the infimum is over all common Stinespring representations
\begin{equation*}%\label{c-S-r}
\Phi(\rho)=\Tr_E V_{\Phi}\rho V^*_{\Phi}\quad\textrm{and}\quad\Psi(\rho)=\Tr_E V_{\Psi}\rho V^*_{\Psi}.
\end{equation*}
It follows from definitions (\ref{d-norm}),(\ref{b-dist+}) and the relations in (\ref{B-d-s-r}) that
\begin{equation*}%\label{DB-rel}
\textstyle\frac{1}{2}\|\Phi-\Psi\|_{\diamond}\leq\beta(\Phi,\Psi)\leq\sqrt{\|\Phi-\Psi\|_{\diamond}}
\end{equation*}
for any channels $\Phi$ and $\Psi$. By representation (\ref{b-dist}) this implies the relations (\ref{star}) which show the selective continuity of the Stinespring representation w.r.t. the diamond-norm topology on the set of quantum channels and the operator-norm topology on the set of Stinespring isometries.

The topology (convergence) generated by the diamond-norm  on the set of infinite-dimensional quantum channels is too
strong for analysis of real variations of such  channels: there are infinite-dimensional channels with  close physical parameters such that the diamond-norm distance between them is equal to $2$ \cite{W-EBN}. In this case it is natural to use the (substantially weaker)  strong convergence (\ref{star+}) of quantum channels studied in detail in \cite{AQC,Wilde+}.

Let $H_A$ be any unbounded densely defined positive (semidefinite) operator on $\H_A$ having discrete spectrum of finite multiplicity and $E_0$ is the minimal eigenvalue of $H_A$. It is shown in \cite{SCT} that the strong convergence of quantum channels is generated by any of the \emph{energy-constrained diamond norms}
\begin{equation}\label{ecd}
\|\Phi\|^E_{\diamond}\doteq \sup_{\rho\in\S(\H_{AR}), \Tr H_A\rho_A\leq E}\|\Phi\otimes \id_R(\rho)\|_1,\quad E>E_0.
\end{equation}
of a Hermitian-preserving superoperator $\Phi:\T(\H_A)\rightarrow\T(\H_B)$, where $R$ is any system. These norms are independently introduced in  \cite{W-EBN}, where a detailed analysis of their properties are presented.\footnote{Slightly different energy-constrained diamond norms are used in \cite{Pir}.} The energy-constrained diamond norms turned out to be a useful tool for quantitative continuity analysis of basic capacities of energy-constrained infinite-dimensional channels \cite{SCT,W-EBN,CID}. These norms are also used in study of quantum dynamical semigroups \cite{W-EBN,Datta,QDS}.

\section{Norms on $\B(\H)$ generating the strong operator topology on bounded subsets of $\B(\H)$.}

In this section we consider norms on $\B(\H)$ generating the strong operator topology on bounded subsets of $\B(\H)$, in particular, on the unit ball of $\B(\H)$.

Let $H$ be any positive (semidefinite) densely defined operator on $\H$ and $E_0=\inf\limits_{\|\varphi\|=1}\langle\varphi|H|\varphi\rangle$. For given $E>E_0$ consider the function on $\B(\H)$ defined as
\begin{equation}\label{ec-on}
 \|A\|_E\doteq \sup_{\rho\in\S(\H),\Tr H\rho\leq E}\sqrt{\Tr A\rho A^*}
\end{equation}
(the supremum is over quantum states $\rho$ satisfying the inequality $\Tr H\rho\leq E$).
\smallskip

\begin{proposition}\label{ec-on-p1}
\emph{The function $A\mapsto\|A\|_E$ defined in (\ref{ec-on}) is a norm on $\B(\H)$. For any given operator  $\,A\in\B(\H)$ the following properties hold:
}\begin{enumerate}[a)]
  \item [\textup{a)}] \emph{$\|A\|_E$ tends to $\|A\|$ as $E\rightarrow+\infty$;}
  \item [\textup{b)}] \emph{the function $E\mapsto\|A\|^2_E$ is concave and nondecreasing on $\,(E_0,+\infty)$;}
  \item [\textup{c)}] \emph{$\|A\varphi\|\leq K_{\varphi}\|A\|_E$ for any unit vector $\varphi$ in $\H$ with finite $E_{\varphi}\doteq\langle\varphi|H|\varphi\rangle$, where $\,K_{\varphi}=1\,$ if
$\,E_{\varphi}\leq E\,$ and $\,K_{\varphi}=\sqrt{(E_{\varphi}-E_0)/(E-E_0)}$ otherwise.}
\end{enumerate}
\end{proposition}

\emph{Proof.} Almost all assertions of the proposition can be easily derived from definition (\ref{ec-on}).

To prove the inequality $\|A+B\|_E\leq \|A\|_E+\|B\|_E$ one should take for given arbitrary $\varepsilon>0$ a state $\rho$ such that
$\|A+B\|_E\leq \sqrt{\Tr|A+B|^2\rho}+\varepsilon$ and $\Tr H\rho\leq E$. Then, by using the spectral decomposition of $\rho$, basic properties of the norm in $\H$ and the Cauchy-Schwarz inequality
it is easy to show that
$$
\sqrt{\Tr|A+B|^2\rho}\leq\sqrt{\Tr|A|^2\rho}+\sqrt{\Tr|B|^2\rho}\leq \|A\|_E+\|B\|_E.
$$

To prove property c) take any unit vector $\varphi\in\H$ with finite $E_{\varphi}$ and arbitrary $\varepsilon>0$. Let
$\rho=(1-K^{-2}_{\varphi})|\phi_{\varepsilon}\rangle\langle\phi_{\varepsilon}|+K^{-2}_{\varphi}|\varphi\rangle\langle\varphi|$,
where $\phi_{\varepsilon}$ is a vector in $\H$ such that $\langle\phi_{\varepsilon}|H|\phi_{\varepsilon}\rangle\leq E_0+\varepsilon$.
Then $\Tr H\rho\leq E+\varepsilon$ and hence
$$
K^{-1}_{\varphi}\|A\varphi\|\leq \sqrt{\Tr A\rho A^*}\leq \|A\|_{E+\varepsilon}.
$$
By passing to the limit $\varepsilon\rightarrow0^+$ we obtain the required inequality. $\square$\smallskip

The norm $\,\|\cdot\|_E$ defined in (\ref{ec-on}) will be called \emph{the operator E-norm}. We will essentially use the following\smallskip

\begin{proposition}\label{ec-on-p2}
\emph{If $\,H$ is  an unbounded densely defined positive operator on $\H$ having discrete spectrum $\{E_k\}_{k\geq0}$ of finite multiplicity and $E>E_0$ then the  operator E-norm $\,\|\cdot\|_E$  generates the strong operator topology on bounded subsets of $\B(\H)$.}
\end{proposition}\smallskip

\emph{Proof.} The set of vectors $\varphi$ in $\H$ with finite $E_{\varphi}\doteq\langle\varphi|H|\varphi\rangle$ is dense in $\H$. So, by using property c) in Proposition \ref{ec-on-p1} it is easy to show the strong convergence of any sequence $\{A_n\}\subset\B(\H)$ to an operator $A_0\in\B(\H)$ provided that $\|A_n-A_0\|_E$ tends to zero as $n\rightarrow+\infty$ and $\sup_n \|A_n\|<+\infty$.

To prove the converse implication note that the assumed properties of the operator $H$ guarantee,  by the Lemma in \cite{H-c-w-c}, the compactness of the
subset $\C_{H,E}$ of $\S(\H)$ determined by the inequality $\,\Tr H\rho\leq E\,$. So, the supremum in definition (\ref{ec-on}) is attained at some state $\rho(A)\in\C_{H,E}$. Assume that $\{A_n\}$ is a sequence in $\B(\H)$ strongly converging to an operator $A_0\in\B(\H)$ such that $\sup_n \|A_n\|=M<+\infty$ and $\|A_n-A_0\|_E$ does not tend to zero as
$n\rightarrow+\infty$. Denote the state $\rho(A_n-A_0)$ by $\rho_n$. By passing to a subsequence we may assume that $\|A_n-A_0\|_E\geq\varepsilon$ for some positive $\varepsilon$ and all $n$ and that
the sequence $\{\rho_n\}$ converges to some state $\rho_0\in\C_{H,E}$ (by the compactness of $\C_{H,E}$). We have
$$
\begin{array}{c}
\|A_n-A_0\|^2_E=\Tr|A_n-A_0|^2\rho_0+\Tr|A_n-A_0|^2(\rho_n-\rho_0)\\\\
\;\qquad\leq \Tr|A_n-A_0|^2\rho_0+ 4M^2\|\rho_n-\rho_0\|_1.
\end{array}
$$
By using the spectral decomposition of $\rho_0$ it is easy to show that the first term in the r.h.s. of this inequality tends to zero as $n\rightarrow+\infty$. This contradicts the above assumption. $\square$\smallskip

The following assertions are easily proved by using definition of the norm $\|\cdot\|_E$. \smallskip

\begin{lemma}\label{E-norm-p}
\noindent A) \emph{For arbitrary operators $A$ and $B$ in $\B(\H)$ the following inequalities hold
$$
m(A)\|B\|_E\leq \|AB\|_E\leq \|A\|\|B\|_E,
$$
where $m(A)$ is the infimum of the spectrum of the operator $|A|=\sqrt{A^*A}$.}\smallskip

\noindent B) \emph{For arbitrary operators $A$ and $B$ in $\B(\H)$ such that $\,\langle A\varphi |B\varphi\rangle=0$ for any $\varphi\in\H$ the following inequalities hold}
$$
\max\!\shs\left\{\|A\|_{E},\|B\|_{E}\shs\right\}\leq \|A+B\|_{E}\leq \sqrt{\left[\|A\|_{E}\right]^2+\left[\|B\|_{E}\right]^2}.
$$
\end{lemma}

\section{The Kretschmann-Schlingemann-Werner theorem and its generalizations.}

The selective continuity of  the Stinespring representation w.r.t. the diamond-norm topology on the set of quantum channels and the operator-norm topology on the set of Stinespring isometries is shown by Kretschmann, Schlingemann and Werner in \cite{Kr&W+,Kr&W}. Theorem 1 in \cite{Kr&W} (stated in the operator algebras settings) implies that
\begin{equation}\label{KSW-th}
\textstyle\frac{1}{2}\|\Phi-\Psi\|_{\diamond}\leq\inf\|V_{\Phi}-V_{\Psi}\|\leq\sqrt{\|\Phi-\Psi\|_{\diamond}}
\end{equation}
for any channels $\Phi$ and $\Psi$, where the infimum is over all common Stinespring representations
\begin{equation}\label{c-S-r}
\Phi(\rho)=\Tr_E V_{\Phi}\rho V^*_{\Phi}\quad\textrm{and}\quad\Psi(\rho)=\Tr_E V_{\Psi}\rho V^*_{\Psi}.
\end{equation}
Inequalities (\ref{KSW-th}) are proved in \cite{Kr&W} by showing that the infimum in (\ref{KSW-th}) coincides with the Bures distance $\beta(\Phi,\Psi)$ defined in (\ref{b-dist+}).

To obtain a generalization of the Kretschmann-Schlingemann-Werner theorem (the KSW-theorem in what follows) consider the \emph{energy-constrained Bures distance}
\begin{equation}\label{ec-b-dist}
\beta_E(\Phi,\Psi)=\sup_{\rho\in\S(\H_{AR}), \Tr H_A\rho_A\leq E} \beta(\Phi\otimes \id_R(\rho),\Psi\otimes \id_R(\rho)), \quad E> E_0,
\end{equation}
between quantum channels $\Phi$  and $\Psi$ from $A$ to $B$  induced by a positive operator $H_A$ on the space $\H_A$ treated as a Hamiltonian of the input system $A$ (here $\,R\,$ is an infinite-dimensional quantum system and $E_0$ is the infimum of the spectrum of $H_A$). This distance is topologically equivalent to the distance induced by the energy-constrained diamond norm (\ref{ecd}), it is introduced in \cite{CID} for quantitative continuity analysis of information characteristics of energy-constrained infinite-dimensional channels. Properties of the energy-constrained Bures distance are presented in Proposition 1 in \cite{CID}. In particular, it is shown in \cite{CID} (by modifying the arguments from the proof of Theorem 1 in \cite{Kr&W}) that
\begin{equation}\label{ec-b-dist+}
\beta_E(\Phi,\Psi)=\inf \sup_{\rho\in\S(\H_{A}),\Tr H_A\rho\leq E}\sqrt{\Tr(V_{\Phi}-V_{\Psi})\rho\shs(V^*_{\Phi}-V^*_{\Psi})},
\end{equation}
where the infimum is over all common Stinespring representations (\ref{c-S-r}), and that this infimum is attainable.

The operator E-norms introduced in Section 3  are obviously generalized to operators between different Hilbert spaces.
By using these  norms and representation (\ref{ec-b-dist+}) one can obtain the following E-version of the KSW-theorem and its modifications.\smallskip

\begin{theorem}\label{KSW-EV} \emph{Let $H_A$ be  an unbounded densely defined positive operator on a Hilbert space $\H_A$ having discrete spectrum $\{E_k\}_{k\geq0}$ of finite multiplicity, $E>E_0$,
$\|\cdot\|^E_{\diamond}$ the energy-constrained diamond norm defined in (\ref{ecd}),
$\beta_E$  the energy-constrained Bures distance defined in (\ref{ec-b-dist}) and $\,\|\cdot\|_E$ the operator E-norm  defined in (\ref{ec-on}) with $H=H_A$.} \smallskip

\noindent A) \emph{For any quantum channels $\Phi$ and $\Psi$ from $A$ to $B$ the following relations hold
\begin{equation}\label{KSW-EV-rel}
\textstyle{\frac{1}{2}}\displaystyle\|\Phi-\Psi\|^E_{\diamond}\leq\beta_{E}(\Phi,\Psi)=\inf_{V_{\Phi},V_{\Psi}}
\|V_{\Phi}-V_{\Psi}\|_E\leq\sqrt{\|\Phi-\Psi\|^E_{\diamond}},
\end{equation}
where the infimum is over all common Stinespring representations (\ref{c-S-r}). This infimum  is attainable.}\smallskip

\noindent B) \emph{For a given quantum channel $\,\Phi$ from $A$ to $B$ there exist a separable Hilbert space $\H_E$ and a Stinespring isometry $V_{\Phi}:\H_A\rightarrow\H_{BE}$ of this channel with the following property: for any  quantum channel $\,\Psi$ from $A$ to $B$ there is a Stinespring isometry $V_{\Psi}:\H_A\rightarrow\H_{BE}$ of $\,\Psi$ such that}
\begin{equation*}%\label{star++}
\|V_{\Phi}-V_{\Psi}\|_E=\beta_{E}(\Phi,\Psi).
\end{equation*}

\noindent C) \emph{If $\,V_{\Phi}:\H_A\rightarrow\H_{BE}$  is the operator from a given Stinespring representation of a quantum channel $\,\Phi$ then
\begin{equation}\label{vl-eq}
\inf_{V_{\Psi}}\|V_{\Phi}-V_{\Psi}\|_E\leq 2\beta_E(\Phi,\Psi)
\end{equation}
for any quantum channel  $\,\Psi$ from $A$ to $B$ with the Choi rank not exceeding $\,\dim\H_{E}$, where the infimum is over all Stinespring representations of $\,\Psi$ with the same environment space $\H_{E}$.}
\end{theorem}\smallskip

\emph{Proof.} A) The equality in ({\ref{KSW-EV-rel}) is the representation (\ref{ec-b-dist+}) rewritten by using the operator E-norms. The inequalities in ({\ref{KSW-EV-rel}) follow from the definitions  (\ref{ecd}) and  (\ref{ec-b-dist}) and the inequalities in (\ref{B-d-s-r}).

B) Assume that  $V_{\Phi}$ is the isometry from any Stinespring representation (\ref{St-rep}) with infinite-dimensional space $\H_E$. Let
$\tilde{V}_{\Phi}$ be the isometry from $\H_A$ into the space
$$
\H_B\otimes(\H^1_E\oplus\H^2_E)=(\H_B\otimes\H^1_E)\oplus(\H_B\otimes\H^2_E),
$$
where $\H^1_E$ and $\H^2_E$ are copies of $\H_E$, defined by setting
$\tilde{V}_{\Phi}|\varphi\rangle=V_{\Phi}|\varphi\rangle\oplus|0\rangle$ for any $\varphi\in\H_A$.

Since any separable Hilbert space can be isometrically embedded into $\H_E$, we may assume that any channel $\Psi$ from $A$ to $B$ has a Stinespring representation with the same environment space $\H_E$. Denote by $V_{\Psi}$ a given Stinespring isometry of the channel $\Psi$ in this representation. The arguments from the proof of Proposition 1 in \cite{CID} (obtained by simple modification of the proof of Theorem 1 in \cite{Kr&W}) show that
\begin{equation}\label{beta-exp}
\beta_E(\Psi,\Phi)=\inf_{C\in\B_1(\H_E)}\|\tilde{V}_{\Psi}^C-\tilde{V}_{\Phi}\|_E=
\|\tilde{V}^{C_0}_{\Psi}-\tilde{V}_{\Phi}\|_E
\end{equation}
for some $C_0\in\B_1(\H_E)$, where $\B_1(\H_E)$ is the unit ball of $\B(\H_E)$ and
$\tilde{V}^C_{\Psi}:\H_A\rightarrow\H_B\otimes(\H^1_E\oplus\H^2_E)$ is
the Stinepring isometry  of the channel $\Psi$ defined by setting
$$
\tilde{V}^C_{\Psi}|\varphi\rangle=(I_B\otimes C)V_{\Psi}|\varphi\rangle\oplus \left(I_B\otimes\sqrt{I_{E}-C^*C}\right)V_{\Psi}|\varphi\rangle
$$
for any $\varphi\in\H_A$ (we assume here that the isometry  $V_{\Psi}$ acts from $\mathcal{H}_A$ to $\mathcal{H}_B\otimes \mathcal{H}^2_E$ and the contraction $C$ acts from $\mathcal{H}^2_E$ to $\mathcal{H}^1_E$).
This implies assertion B of the theorem  with the isometry $\tilde{V}_{\Phi}$ in the role of $V_{\Phi}$.

C) Since any $k$-dimensional Hilbert space can be isometrically embedded into $\H_E$ provided that $k\leq \dim\H_E$, we may assume that any channel $\Psi$ with the Choi rank not exceeding $\dim\H_E$ has a Stinespring representation with the same environment space $\H_E$. By repeating the arguments from the proof of assertion B we construct the operators $\tilde{V}_{\Phi}$ and $\tilde{V}^{C_0}_{\Psi}$. Assume first that the operator $C_0$ is nondegenerate, i.e. $\ker C_0=\{0\}$. Let $U$ be the isometry from the polar decomposition of $C_0$, i.e. $C_0=U|C_0|$. Since $\,\|\tilde{V}^{C_0}_{\Psi}-\tilde{V}_{\Phi}\|_E=\beta_E(\Psi,\Phi)$, it follows from Lemma \ref{E-norm-p}B  that
\begin{equation}\label{s-ub}
\!\!\|(I_B\otimes C_0)V_{\Psi}-V_{\Phi}\|_E\leq\beta_E(\Psi,\Phi)\quad\textrm{and}\quad \left\|\left(I_B\otimes\sqrt{I_{E}-|C_0|^2}\right)\!V_{\Psi}\right\|_E\!\leq\beta_E(\Psi,\Phi)\!
\end{equation}
Hence the triangle inequality and Lemma \ref{E-norm-p}A imply that
\begin{equation}\label{s-ub+}
\begin{array}{c}
\|(I_B\otimes U)V_{\Psi}-V_{\Phi}\|_E\leq \|(I_B\otimes C_0)V_{\Psi}-V_{\Phi}\|_E\\\\+\|(I_B\otimes C_0)V_{\Psi}-(I_B\otimes U)V_{\Psi}\|_E
\leq \beta_E(\Psi,\Phi)+\|I_B\otimes (I_E-|C_0|)V_{\Psi}\|_E.
\end{array}
\end{equation}
Since $C_0$ is a contraction, by using Lemma \ref{E-norm-p}A and the second inequality in (\ref{s-ub}) we obtain
$$
\|I_B\otimes (I_E-|C_0|)V_{\Psi}\|_E\leq\|I_B\otimes (I_E-|C_0|^2)V_{\Psi}\|_E\leq\|I_B\otimes \sqrt{I_E-|C_0|^2}V_{\Psi}\|_E\leq\beta_E(\Psi,\Phi).
$$
Thus, it follows from (\ref{s-ub+}) that $\|(I_B\otimes U)V_{\Psi}-V_{\Phi}\|_E\leq2\beta_E(\Psi,\Phi)$. Since $U$ is an isometry,
$(I_B\otimes U)V_{\Psi}$ is a Stinespring isometry for $\Psi$.\smallskip

To omit the assumption $\ker C_0=\{0\}$ it suffices to show that the infimum in (\ref{beta-exp})
can be taken over the subset $\B^\mathrm{n}_1(\H_E)$ of $\B_1(\H_E)$ consisting of nondegenerate operators. Since
$\B^\mathrm{n}_1(\H_E)$ is dense in $\B_1(\H_E)$ in the weak operator topology, this can be easily done by noting that
$$
\|\tilde{V}^{C}_{\Psi}-\tilde{V}_{\Phi}\|_E
=\sup_{\rho\in\S(\H_{A}),\Tr H_A\rho\leq E}\sqrt{2-2\Re\shs\Tr V_{\Phi}^*(I_B\otimes C)V_{\Psi}\rho}
$$
and by using compactness of the set of states $\rho$ in $\S(\H_{A})$ satisfying the inequality $\Tr H_A\rho\leq E$ \cite{H-c-w-c}.
$\square$
\smallskip

\begin{remark}\label{main-r+}  The assumption that $H_A$ is an unbounded positive operator having discrete spectrum of finite multiplicity is used only in the proof of part C of Theorem \ref{KSW-EV}. So, parts A and B of this theorem are valid for any positive densely defined operator $H_A$ and $E>E_0=\inf\limits_{\|\varphi\|=1}\langle\varphi|H_A|\varphi\rangle$.
If $H_A=I_A$ and $E>1$ then part A of Theorem \ref{KSW-EV} is the standard KSW-theorem while part B is its strengthened version adapted
for analysis of uniformly converging sequences of quantum channels (see Remark \ref{main-r} below).
\end{remark}

\section{Characterization of the strong convergence in terms of the Stinespring and Kraus representations.}

Theorem \ref{KSW-EV} in Section 4, Proposition 3 in \cite{SCT} and Proposition \ref{ec-on-p2} in Section 3 imply the following  characterisation of the strong convergence of quantum channels in terms of their Stinespring's  representations.\smallskip

\begin{theorem}\label{main-t} \emph{Let $H_A$ be  an unbounded densely defined positive operator on a Hilbert space $\H_A$ having discrete spectrum $\{E_k\}_{k\geq0}$ of finite multiplicity, $E>E_0$, $\|\cdot\|^E_{\diamond}$ the energy-constrained diamond norm defined in (\ref{ecd}), $\beta_E$ the energy-constrained Bures distance defined in (\ref{ec-b-dist}) and $\|\cdot\|_E$ the operator E-norm  defined in (\ref{ec-on}) with $H=H_A$.}
\smallskip

\noindent A) \emph{If a sequence of isometries $V_n:\H_A\rightarrow\H_{BE}\doteq\H_B\otimes\H_E$ strongly converges to an isometry $V_0:\H_A\rightarrow\H_{BE}$ then the sequence of channels $\,\Phi_n(\rho)=\Tr_E V_n\rho V^*_n\,$ strongly converges to the channel $\,\Phi_0(\rho)=\Tr_E V_0\rho V^*_0\,$ and}
$$
\textstyle{\frac{1}{2}}\|\Phi_n-\Phi_0\|^E_{\diamond}\leq\beta_{E}(\Phi_n,\Phi_0)\leq\|V_{n}-V_{0}\|_E\qquad \forall n.
$$

\noindent B) \emph{If  a sequence of quantum channels $\,\Phi_n:A\rightarrow B$ strongly converges to a quantum channel $\,\Phi_0:A\rightarrow B$ then there exist a separable Hilbert space $\H_E$ and a sequence of isometries $V_n:\H_A\rightarrow\H_{BE}$ strongly converging to an isometry $V_0:\H_A\rightarrow\H_{BE}$ such that $\,\Phi_n(\rho)=\Tr_E V_n\rho V^*_n\,$ for all $\,n\geq0$ and}
\begin{equation*}%\label{star++}
\|V_{n}-V_{0}\|_E=\beta_{E}(\Phi_n,\Phi_0)\leq \sqrt{\|\Phi_n-\Phi_0\|^E_{\diamond}}\qquad \forall n.
\end{equation*}
\emph{If  $\,V_0:\H_A\rightarrow\H_{BE_0}$ is a given Stinespring isometry for $\,\Phi_0$, where $\H_{E_0}$ is an infinite-dimensional Hilbert space,  then for any $\varepsilon>0$ there exists  a sequence of isometries $V_n:\H_A\rightarrow\H_{BE_0}$ strongly converging to the isometry $V_0$ such that $\,\Phi_n(\rho)=\Tr_{E_0} V_n\rho V^*_n\,$ and
\begin{equation}\label{2-est}
\|V_{n}-V_{0}\|_E\leq 2\beta_{E}(\Phi_n,\Phi_0)+\varepsilon\leq 2\sqrt{\|\Phi_n-\Phi_0\|^E_{\diamond}}+\varepsilon\qquad \forall n.
\end{equation}
If the Choi rank of all the channels $\,\Phi_n$ does not exceed $\,m<+\infty\,$ then the above assertion is valid provided that $\,\dim\H_{E_0}\geq m$.}
\end{theorem}\smallskip

Factor $"2"$ and the arbitrarily small summand $\varepsilon$ in (\ref{2-est}) is a cost of the possibility to take the sequence $\{V_n\}$ of Stinespring isometries representing the sequence $\{\Phi_n\}$
that strongly converges to a \emph{given} Stinespring isometry $V_0:\H_A\rightarrow\H_{BE_0}$ of the channel $\Phi_0$.\smallskip

\emph{Proof.} Assertion A follows directly from
the first inequality in (\ref{KSW-EV-rel}), Proposition 3B in \cite{SCT} and Proposition \ref{ec-on-p2} in Section 3.

To prove B note that for any sequence of quantum channels $\Phi_n:A\rightarrow B$ strongly converging to a channel $\Phi_0:A\rightarrow B$  Theorem \ref{KSW-EV}B implies existence of a separable Hilbert space $\H_E$, a sequence of isometries $V_n:\H_A\rightarrow \H_{BE}$ and an isometry $V_0:\H_A\rightarrow\H_{BE}$ such that $\,\Phi_n(\rho)=\Tr_E V_n\rho V^*_n\,$ and $\,\|V_{n}-V_{0}\|_E=\beta_{E}(\Phi_n,\Phi_0)\,$ for all $n\geq0$. So, the second inequality in (\ref{KSW-EV-rel}), Proposition 3B in \cite{SCT} and Proposition \ref{ec-on-p2} in Section 3 show the strong convergence of the sequence $\{V_n\}$ to the isometry $V_0$.

The last assertion of B is proved similarly by using Theorem \ref{KSW-EV}C.
$\square$
\smallskip

The Stinespring representation (\ref{St-rep}) implies that any quantum channel $\Phi:A\rightarrow B$ has the Kraus representation
$$
\Phi(\rho)=\sum_{i=1}^m A_i\rho A_i^*,
$$
where $\{A_i\}_{i=1}^m$ is a collection of linear operators from $\H_{A}$ to $\H_{B}$ such that $\sum_{i=1}^m A^*_i A_i=I_A$ and $m=\dim\H_E\leq+\infty$ \cite{H-SCI,Wilde,Kraus}. Theorem \ref{main-t} implies the following necessary and sufficient conditions of the strong convergence of quantum channels in terms of their Kraus representations.\smallskip

\begin{corollary}\label{main-c+} A) \emph{Let $\{\{A_i^n\}_n\}_{i=1}^m$ be a set of $m\leq+\infty$ sequences of linear operators from $\H_{A}$ to $\H_{B}$ such that $\,s\shs\textup{-}\lim\limits_{n\rightarrow\infty} A_i^n=A_i^0$ for each $\,i=\overline{1,m}$ and $\sum_{i=1}^m [A_i^n]^*A_i^n=I_{A}$ for all $n\geq0$. The sequence of
channels $\,\Phi_n(\rho)=\sum_{i=1}^m A_i^n\rho [A_i^n]^*$ strongly converges to the channel $\,\Phi_0(\rho)=\sum_{i=1}^m A_i^0\rho [A_i^0]^*$.}\smallskip

\noindent B) \emph{Let $\,\{\Phi_n\}$ be a sequence of quantum channels from $A$ to $B$ strongly converging to a quantum channel $\,\Phi_0$
and $\,m$ the maximal Choi rank of all the channels $\,\Phi_n$, $n\geq0$, if it is finite and $m=+\infty$ otherwise. There exists
a set  $\{\{A_i^n\}_n\}_{i=1}^m$ of $m$ sequences of linear operators from $\H_{A}$ to $\H_{B}$ such that $\Phi_n(\rho)=\sum_{i=1}^m A_i^n\rho [A_i^n]^*$ for all $n\geq0$ and $\,s\shs\textup{-}\lim\limits_{n\rightarrow\infty} A_i^n=A_i^0$ for each $\,i=\overline{1,m}$.}\smallskip

\emph{If $\,\Phi_0(\rho)=\sum_{i=1}^{m_0} A_i^0\rho [A_i^0]^*$, $m_0\in [m,+\infty]$, is a given Kraus representation of $\Phi_0$ then there exists
a set $\{\{A_i^n\}_n\}_{i=1}^{m_0}$ of $m_0$ sequences of linear operators from $\H_{A}$ to $\H_{B}$ such that $\,\Phi_n(\rho)=\sum_{i=1}^{m_0} A_i^n\rho [A_i^n]^*$ for all $\,n$ and $\,s\shs\textup{-}\lim\limits_{n\rightarrow\infty} A_i^n=A_i^0$ for each $\,i=\overline{1,m_0}$.}\smallskip
\end{corollary}

\emph{Proof.} A) Assume that $\{\tau_i\}_{i=1}^{m}$ is a basic in a $m$-dimensional Hilbert space $\H_E$. Then  $\Phi_n(\rho)=\Tr_E V_n\rho V^*_n$ for all $n\geq0$, where $\{V_n\}$ is a sequence of operators
defined by setting  $V_n|\varphi\rangle=\sum_{i=1}^{m} A^n_i|\varphi\rangle\otimes|\tau_i\rangle$ for any $\varphi\in \H_A$. Since $\,s\shs\textup{-}\lim_{n\rightarrow\infty} A^n_i=A^0_i$ for all $i=\overline{1,m}$, the sequence $\{V_n|\varphi\rangle\}$ weakly converges to the vector $V_0|\varphi\rangle$. The weak convergence of this sequence implies its  convergence in the norm of $\H_{BE}$, since all the operators $V_n$, $n\geq0$, are isometries (this follows from the condition $\sum_{i=1}^{m} [A^n_i]^* A^n_i=I_A$ for all $n$). Thus, Theorem \ref{main-t}A implies strong convergence
the sequence $\,\{\Phi_n\}$ to the channel $\,\Phi_0$.
\smallskip

B) It suffices to prove the last assertion of B. Let $\{\tau_i\}_{i=1}^{m_0}$ be a basic in $m_0$-dimensional Hilbert space $\H_{E_0}$ and $V_0$ the isometry
from $\H_A$ to $\H_{BE_0}$ defined by setting  $V_n|\varphi\rangle=\sum_{i=1}^{m_0} A^0_i|\varphi\rangle\otimes|\tau_i\rangle$ for any $\varphi\in \H_A$.
Theorem \ref{main-t}B implies
existence of a sequence $\{V_n\}$ of isometries from $\,\H_{A}$ into $\H_{BE_0}$ such that $\Phi_n(\rho)=\Tr_E V_n\rho V^*_n$ for all $\,n$ and $\,s\shs\textup{-}\lim_{n\rightarrow\infty} V_n=V_0$. Let $A^n_i$ be the operator from $\H_A$ to $\H_B$ such that $\langle\psi|A^n_i|\varphi\rangle=\langle\psi\otimes \tau_i|V_n|\varphi\rangle$ for any $\varphi\in\H_A$ and $\psi\in\H_B$. Then $\Phi_n(\rho)=\sum_{i=1}^{m_0} A^n_i\rho [A^n_i]^*$ for all $n$. By noting that $V_n|\varphi\rangle=\sum_{i=1}^{m_0} A^n_i|\varphi\rangle\otimes|\tau_i\rangle$ for any $\varphi\in\H_A$ and $n$ it is easy to show that
$\,s\shs\textup{-}\lim_{n\rightarrow\infty} A^n_i=A^0_i$ for each $i$. $\square$
\smallskip

If a quantum channel $\Phi:A\rightarrow B$ has Stinespring representation (\ref{St-rep}) then the quantum  channel
\begin{equation}\label{c-channel}
\mathfrak{T}(\mathcal{H}_A)\ni
\rho\mapsto\widehat{\Phi}(\rho)=\mathrm{Tr}_{B}V_{\Phi}\rho
V_{\Phi}^{*}\in\mathfrak{T}(\mathcal{H}_E)
\end{equation}
is called \emph{complementary} to the channel $\Phi$
\cite[Ch.6]{H-SCI}. The complementary channel is uniquely defined up to \emph{isometrical equivalence}, i.e.
if $\,\widehat{\Phi}':A\rightarrow E'$ is the channel defined by formula  (\ref{c-channel})
via some other Stinespring isometry  $\,V'_{\Phi}:\H_A\rightarrow\H_B\otimes\H_{E'}\,$ then there exists a partial isometry
$W:\H_E\rightarrow\H_{E'}$ such that $\widehat{\Phi}'(\rho)=W\widehat{\Phi}(\rho)W^*$ and $\widehat{\Phi}(\rho)=W^*\widehat{\Phi}'(\rho)W$ for all $\rho\in \S(\H_A)$ \cite{H-c-ch}.

Let $H_A$ be  a positive operator on $\H_A$, $E>E_0$ and $\beta_E$ the corresponding energy-constrained Bures distance defined in (\ref{ec-b-dist}). It follows from representation (\ref{ec-b-dist+}) that for any quantum channels $\Phi$ and $\Psi$ from $A$ to $B$ one can find complementary channels $\widehat{\Phi}$ and $\widehat{\Psi}$  from $A$ to some system $E$ such that\footnote{Since a complementary channel is defined up to the isometrical equivalence the quantity $\beta_E(\widehat{\Phi},\widehat{\Psi})$ depends on concrete realizations of the complementary channels  $\widehat{\Phi}$ and $\widehat{\Psi}$.}
\begin{equation*}%\label{ec-b-dist+}
\beta_E(\widehat{\Phi},\widehat{\Psi})\leq\beta_E(\Phi,\Psi).
\end{equation*}

Theorem \ref{main-t}B implies the following observations which show (due to Proposition 1 in \cite{CID}) the \emph{uniform selective continuity of the complementary operation} $\,\Phi\mapsto\widehat{\Phi}\,$ with respect to the strong convergence of quantum channels.\smallskip

\begin{corollary}\label{main-c} \emph{Let $\,\{\Phi_n\}$ be a sequence of quantum channels from $A$ to $B$ strongly converging to a quantum channel $\,\Phi_0$ and $\,m$ the maximal Choi rank of all the channels $\,\Phi_n$ if it is finite and $m=+\infty$ otherwise.}\smallskip

A) \emph{There exists a sequence $\,\{\mathrm{\Psi}_n\}$ of channels  from $A$ to some system $E$ strongly converging  to a channel $\,\mathrm{\Psi}_0$ such that $\,\mathrm{\Psi}_n=\widehat{\mathrm{\Phi}}_n$ and $\,\beta_E(\mathrm{\Psi}_n,\mathrm{\Psi}_0)\leq\beta_E(\mathrm{\Phi}_n,\mathrm{\Phi}_0)$ for all $\,n\geq0$.}\smallskip

B) \emph{If $\,\widehat{\Phi}_0$ is a given complementary channel to the channel $\Phi_0$ acting from $A$ to a system $E_0$ such that $\dim\H_{E_0}\geq m$ then for any $\varepsilon>0$ there exists a sequence $\,\{\Psi_n\}$ of channels  from $A$ to $E_0$ strongly converging  to the channel $\,\widehat{\Phi}_0$ such that $\,\Psi_n=\widehat{\Phi}_n$ and}
\begin{equation*}%\label{c-ineq-vl}
\beta_E(\Psi_n,\widehat{\Phi}_0)\leq2\beta_E(\Phi_n,\Phi_0)+\varepsilon \qquad \forall n.
\end{equation*}
\end{corollary}\medskip

\begin{remark}\label{main-r}  Assertion B of Theorem \ref{KSW-EV} is valid with the energy-constrained Bures distance $\beta_E$ and the operator E-norm $\|\cdot\|_E$ replaced, respectively, by the (unconstrained) Bures distance $\beta$ and the operator norm $\|\cdot\|$ (see Remark \ref{main-r+}). It can be used to obtain the versions of Theorem \ref{main-t}B and Corollary \ref{main-c}A in which the strong convergences of channels and operators are replaced by the diamond norm and the operator norm convergences.
\end{remark}

\section{On continuity of the unitary dilation}

By using the Stinespring representation (\ref{St-rep}) it is easy to show  that any quantum channel $\Phi$ from $A$ to $B=A$ can be represented as
\begin{equation}\label{ud-1}
  \Phi(\rho)=\Tr_E\shs U_{\Phi} \rho\otimes\sigma_{0} U_{\Phi}^*,
\end{equation}
where $\sigma_{0}$ is a pure state in $\S(\H_E)$ and $U_{\Phi}$ is a unitary operator on $\H_{AE}$. Representation
(\ref{ud-1}) allows to consider any channel from a quantum system $A$ to itself as a reduction of some  unitary (reversible) evolution of the larger quantum system $AE$ \cite{H-SCI,Wilde,Kraus}.

In general case,  for a given quantum channel $\Phi$ from $A$ to $B$ having representation (\ref{St-rep}) one can find such quantum systems $D$ and $E'$ that
\begin{equation}\label{ud-2}
  \Phi(\rho)=\Tr_{E'}\shs U_{\Phi\shs} \rho\otimes\sigma_{0} U_{\Phi}^*,
\end{equation}
where $\sigma_{0}$ is a pure state in $\S(\H_{D})$ and $U_{\Phi}$ is a unitary operator from $\H_{AD}$ onto $\H_{BE'}$ \cite{H-SCI}. In particular, one can take $D=BE$ and $E'=AE$. For an infinite-dimensional quantum channel $\Phi$ with  representation (\ref{St-rep}) such that $\dim(\H_{BE}\ominus\Ran V_{\Phi})=+\infty$ one can always take $E'=E$. This follows from the fact that any partial isometry $W$ such that $\dim\ker W=\dim\ker W^*=+\infty$ can be extended to a unitary operator \cite{R&S}.

Representations (\ref{ud-1}) and (\ref{ud-2}) are called \emph{unitary dilations} of a quantum channel $\Phi$ \cite{H-SCI,UD+}. Since
(\ref{ud-1}) is a partial case of (\ref{ud-2}), the latter can be called \emph{universal unitary dilation}. In this section we explore
selective continuity of the multi-valued map $\Phi\mapsto U_{\Phi}$ w.r.t. different topologies on the sets of quantum channels and unitary operators.

\subsection{Continuity of the unitary dilation w.r.t. the uniform convergence}

We show first that the KSW-theorem implies selective continuity of the map $\Phi\mapsto U_{\Phi}$ w.r.t. the diamond norm metric on the sets of quantum channels and the operator norm metric on the set of unitary operators.

\begin{proposition}\label{d-n-ud}
\emph{For an arbitrary  sequence $\,\{\Phi_n\}$ of quantum channels from $A$ to $B$ diamond norm converging to a channel $\,\Phi_0$  there  exist quantum systems $D$ and $E$, a sequence $\{U_n\}$ of unitary operators from  $\H_{AD}$ onto $\H_{BE}$  norm converging to a unitary  operator $\,U_0:\H_{AD}\rightarrow\H_{BE}$ and a pure state  $\sigma_0$  in $\S(\H_{D})$ such that
}$$
\Phi_n(\rho)=\Tr_{E\shs} U_n\rho\otimes\sigma_0 U^*_n\quad\textit{ for all }\quad n\geq0.
$$
\end{proposition}

\emph{Proof.} By using the arguments from the proof of Theorem 1 in \cite{Kr&W} one can show existence of a quantum system $E$ and a sequence $\{V_n\}$ of isometries from $\,\H_{A}$ into $\H_{BE}$ norm converging to an isometry $V_0:\H_{A}\rightarrow\H_{BE}$ such that $\Phi_n(\rho)=\Tr_E V_n\rho V^*_n$ for all $\,n\geq0$ (see Remark \ref{main-r}).

Let $C$ and $D$ be infinite-dimensional quantum systems and  $\sigma_0=|\tau_0\rangle\langle\tau_0|$, where $\tau_0$ is any unit vector in $\H_D$. If we identify the space $\H_A$ with the subspace $\H_A\otimes \{c\tau_0\}$ of $\H_{AD}$, then $\{V_n\}$ is a sequence of partial isometries from  $\H_{AD}$ to $\H_{BEC}\cong\H_{AD}$ norm converging to the partial isometry $V_0$ such that $V^*_nV_n=V^*_0V_0$ and  $\dim\ker V^*_nV_n=\dim\ker V_nV^*_n=+\infty$ for all $n\geq 0$. So, the existence of the sequence $\{U_n\}$ with the required properties (with the system $EC$ in the role of $E$) follows from Lemma \ref{udc} below. $\square$

\begin{lemma}\label{udc} \emph{Let $\,\{V_n\}$ be a sequence of partial isometries on a separable Hilbert space $\H$ norm
converging to a partial isometry $V_0$ such that $V_n^*V_n=V_0^*V_0=P$  and $\,\dim\mathrm{Ker}P=\dim\mathrm{Ker}Q_n\leq+\infty,\,$ where $Q_n=V_nV^*_n$,  for all $\,n\geq 0$.  Let $U_0$ be a given unitary operator such that $\,U_0P=V_0$.\footnote{The existence of such operator is guaranteed by the condition $\,\dim\mathrm{Ker}P=\dim\mathrm{Ker}Q_0$.} Then there exists a sequence  $\,\{U_n\}$ of unitary operators  norm converging to the operator $\,U_0$  such that $\,U_nP=V_n$ for all $\,n$.}
\end{lemma}\smallskip

\emph{Proof.}  Since all the partial isometries have the same initial space, the sequence $\{W_n=V_nV_0^*\}$ consists of partial isometries and norm converges to the projector $Q_0=V_0V_0^*$. Note that $W_nW_n^*=Q_n$ and $W^*_nW_n=Q_0$ for all $n$. Assume that $\{\bar{W}_n\}$ is a sequence
of partial isometries norm converging to the projector $R_0=I_{\H}-Q_0$ such that $\bar{W}_n\bar{W}_n^*=R_n\doteq I_{\H}-Q_n$ and $\bar{W}^*_n\bar{W}_n=R_0$ for all $\,n$. Then the sequence of unitary operators $(W_n+\bar{W}_n)U_0$ has the required property.

The sequence $\{\bar{W}_n\}$ can be constructed as follows. Let $T_n=R_nR_0$ and $|T_n|=\sqrt{R_0R_nR_0}$. Since
the sequence  $\{R_n\}$ norm converges to the projector $R_0$, we may assume that $\,\|R_n-R_0\|<1\,$ for all $n$. It is easy to see that the last inequality implies that $\Ran T_n=R_n(\H)$ and $\Ran |T_n|=R_0(\H)$. Let $\bar{W}_n$ be the partial isometry from the polar decomposition of $T_n$, i.e. $T_n=\bar{W}_n|T_n|$, such that $\Ran \bar{W}^*_n=R_0(\H)$. Since the sequences $\{T_n\}$ and $\{|T_n|\}$ norm converges  to the projector $R_0$ \cite{R&S}, it is easy to show that the sequence $\{\bar{W}_n\}$ norm converges to the projector $R_0$ as well.
$\square$

\subsection{Discontinuity of the unitary dilation  w.r.t. the strong convergence}

In this section we show that the map $\Phi\mapsto U_{\Phi}$ is discontinuous in the following sense: there is a sequence $\{\Phi_n\}$ of quantum channels  strongly converging to a channel $\Phi_0$ that can not be represented in the form (\ref{ud-2}) with a sequence $\{U_{\Phi_n}\}$ of unitary operators strongly converging to a unitary operator $U_{\Phi_0}$. Moreover, this discontinuity can not be eliminated by making the state $\sigma_0$ in (\ref{ud-2})  dependent on a channel $\Phi$.

We will use the following observation in which $\Phi^*$ denotes the dual map to a channel $\Phi$ defined by the relation $\Tr\shs \Phi(\rho)B=\Tr\shs\Phi^*\!(B)\rho$ for any $\rho\in\S(\H_A)$, $B\in\B(\H_B)$. The map  $\Phi^*$ is a quantum channel in the Heisenberg picture \cite{H-SCI}.\smallskip

\begin{proposition}\label{ud-nc} \emph{Let $\{U_n\}$ be a sequence of unitary operators from  $\H_{AD}$ onto $\H_{BE}$ converging to a unitary  operator $\,U_0$ in the strong operator topology and $\{\sigma_n\}$ a sequence of states in $\S(\H_{D})$ converging to a state $\sigma_0$. Let $\,\Phi_n(\rho)=\Tr_{E\shs} U_n\rho\otimes\sigma_n U^*_n\,$ be a channel from $A$ to $B$ for any $\,n\geq0$. Then the sequence $\{\Phi_n^*(B)\}$ converges to the operator $\,\Phi_0^*(B)$ in the strong operator topology for any $B\in\B(\H_B)$.}
\end{proposition}\smallskip

The condition $\lim_{n\rightarrow\infty} \sigma_n=\sigma_0$ is necessary and sufficient for  the strong convergence of the channels $\rho\mapsto U_n\rho\otimes\sigma_n U^*_n$ to the channel $\rho\mapsto U_0\rho\otimes\sigma_0 U^*_0$ (provided that $\,s\shs\textup{-}\lim_{n\rightarrow\infty} U_n=U_0$).\smallskip

\emph{Proof.} Since for any converging sequence of states in $\S(\H_{D})$ there is a converging sequence of purifications in
$\S(\H_{DR})$, where $R$ is some system, and $\,s\shs\textup{-}\lim_{n\rightarrow\infty} U_n=U_0$ implies $\,s\shs\textup{-}\lim_{n\rightarrow\infty} U_n\otimes I_R=U_0\otimes I_R$, we may assume that the sequence $\{\sigma_n\}$ consists of pure states. It is easy to see that
$\Phi^*_n(B)=\Tr_{D}[I_A\otimes\sigma_n] [U^*_n B\otimes I_{E\shs} U_n]$. So, the assumed purity of the state $\sigma_n$ implies that
$$
T_n\doteq[I_A\otimes\sigma_n] [U^*_n B\otimes I_{E\,} U_n] [I_A\otimes\sigma_n] = \Phi^*_n(B)\otimes\sigma_n, \quad n\geq 0.
$$
Since the sequence  $\{U^*_n\}$ strongly converges to the operator $U^*_0$,\footnote{Here and in what follows we use the continuity of the map $A\mapsto A^*$ in the strong operator topology on the set of unitary operators \cite{R&S}.} the sequence $\{T_n\}$ strongly converges to the operator $T_0$. It follows that the sequence $\{\Phi_n^*(B)\}$ strongly converges to the operator $\Phi_0^*(B)$ as well. $\square$\smallskip

\begin{corollary}\label{ud-cr-c}
\emph{There exists a  sequence $\{\Phi_n\}$ of quantum channels from an infinite-dimensional quantum system to itself strongly converging to a channel $\,\Phi_0$ such that the channels $\,\Phi_n$ can not be represented in  the form $\,\Phi_n(\rho)=\Tr_{E}\shs U_{n\shs} \rho\otimes\sigma_{n} U_{n}^*$ for all $n\geq0$, where $\{U_{n}\}$ is a sequence of unitary operators from $\H_{AD}$ onto $\H_{BE}$ strongly converging to a unitary operator $\,U_{0}$ and $\,\{\sigma_{n}\}$ is a  sequence  of states in $\S(\H_D)$ converging to a state $\sigma_{0}$.}
\end{corollary}
\smallskip

\emph{Proof.} Let $\H_A=\H_B$ be a separable Hilbert space and $\H_0$ an infinite-dimensional subspace of $\H_A$. Let $\{\tau_i\}_{i\in\mathbb{N}}$ be an orthonormal basic in $\H_0$ and $\psi$  any unit vector in $\H_0^{\bot}$. For each $n$ consider the partial isometry
$$
V_n=\sum_{i\neq n}|\tau_i\rangle\langle \tau_i|+|\psi\rangle\langle \tau_n|.
$$
Then $V^*_n=\sum_{i\neq n}|\tau_i\rangle\langle \tau_i|+|\tau_n\rangle\langle \psi|$ and hence $V^*_nV_n=P_0$,
where $\,P_0=\sum_{i}|\tau_i\rangle\langle \tau_i|\,$ is the projector on the subspace $\H_0$. It is easy to see that
$$
s\shs\textup{-}\lim_{n\rightarrow\infty} V_n=P_0,
$$
while the sequence $\{V^*_n\}$ has no limit in the strong operator topology.

The sequence of the channels $\Phi_n(\rho)=V_n\rho V^*_n+ \bar{P}_0\rho\bar{P}_0$ strongly converges to the channel $\Phi_0(\rho)=P_0\rho P_0+ \bar{P}_0\rho\bar{P}_0$, where $\bar{P}_0=I_{A}-P_0$. It is easy to see that the sequence $\{\Phi^*_n(|\psi \rangle\langle \tau_1|)\}$ does not converge to the operator $\Phi^*_0(|\psi \rangle\langle \tau_1|)=0\,$
in the strong operator topology. $\square$
\smallskip

This result can be treated as discontinuity of the unitary dilation w.r.t. the strong convergence topology on the set of quantum channels and the strong operator topology on the set of unitary operators. Mathematically, this discontinuity is connected with the discontinuity of the map $A\mapsto A^*$ in the strong operator topology on $\B(\H)$.

The discovered discontinuity of the unitary dilation with respect to the strong convergence
has an interesting physical implication described below.

It is known that all the basic topologies on the algebra $\B(\H)$ excepting the norm topology coincide on the set of unitary operators \cite{B&R}. So, the strong convergence topology on the set of infinite-dimensional unitary channels, i.e. channels of the form $\Phi(\rho)=U\rho U^*$, where $U$ is a unitary operator, seems the only reasonable topology on the set of such channels. Therefore, Corollary \ref{ud-cr-c} states the existence of  strongly converging
sequences of quantum channels that \emph{has no prototype} within the standard interpretation
of a channel as a reduced unitary evolution of some larger system (or unitary transformation between larger systems).
This shows that the strong convergence of quantum channels
is\emph{ too weak} for describing physical perturbations of quantum channels. The weakest type of convergence of quantum channels with respect to which the unitary dilation is continuous (in the above sense) is considered in \cite{SSC}.

\section{Some applications}

The representation of strongly converging sequences of quantum channels given by Theorem \ref{main-t}B is a useful tool for continuity analysis  of informational characteristics of quantum channels  w.r.t. the strong convergence.

\subsection{Lower semicontinuity of the entropic disturbance w.r.t. the strong convergence of quantum channels}

A finite or countable collection $\{\rho _{i}\}$ of quantum states with a
probability distribution $\{p_{i}\}$ is called an \textit{ensemble} and
denoted by $\{p_{i},\rho _{i}\}$. The state $\bar{\rho}\doteq \sum_{i}p_{i}\rho _{i}$ is called the \emph{average state} of the ensemble.
  We will also use the notion of \textit{generalized ensemble} as a
Borel probability measure on the set of quantum states, so that previously
defined ensembles correspond to discrete probability measures. We denote by $%
\mathcal{P}(\mathcal{H})$ the set of all Borel probability measures on $%
\mathfrak{S}(\mathcal{H})$ equipped with the topology of weak convergence
\cite{Bil,Par,H-Sh-2}. The set $\mathcal{P}(\mathcal{H})$ can be considered as a complete
separable metric space \cite{Par}. The average state of a generalized
ensemble $\mu \in \mathcal{P}(\mathcal{H})$ is the barycenter of the measure
$\mu $ defined by the Bochner integral
\begin{equation*}
\bar{\rho}(\mu )=\int_{\mathfrak{S}(\mathcal{H})}\rho \mu (d\rho ).
\end{equation*}

The \emph{Holevo quantity} of  a discrete ensemble $\,\{p_{i},\rho _{i}\}$ is defined as
\begin{equation}
\chi (\{p_{i},\rho _{i}\})=\sum_{i}p_{i}H(\rho _{i}\Vert \bar{\rho})=H(\bar{\rho})-\sum_{i}p_{i}H(\rho _{i}),  \label{chi-q-d}
\end{equation}%
where $H(\cdot)$ is the von Neumann entropy and $H(\cdot\|\shs \cdot )$ is the quantum relative entropy (defined in Section 2). The second formula in (\ref{chi-q-d}) is valid under the condition  $H(\bar{\rho})<+\infty $. This quantity is an upper bound on the classical information which can be
obtained by recognizing the states of an ensemble by quantum measurements \cite{H-73}.

The Holevo quantity of a
generalized ensemble $\mu \in \mathcal{P}(\mathcal{H})$ is defined as
\begin{equation*}
\chi (\mu )=\int H(\rho\shs \|\shs \bar{\rho}(\mu ))\mu (d\rho )=H(\bar{\rho}(\mu
))-\int H(\rho )\mu (d\rho ),  %\label{chi-q-d+}
\end{equation*}
where the second formula is valid under the condition $H(\bar{\rho}(\mu))<+\infty $ \cite{H-Sh-2}.

For an ensemble $\mu \in \mathcal{P}(\mathcal{H}_{A})$ its image $\Phi (\mu)$
under a quantum channel $\Phi :A\rightarrow B\,$ is defined as the
ensemble in $\mathcal{P}(\mathcal{H}_{B})$ corresponding to the measure $\mu
\circ \Phi ^{-1}$ on $\mathfrak{S}(\mathcal{H}_{B})$, i.e. $\,\Phi (\mu )[%
\mathfrak{S}_{B}]=\mu[\Phi ^{-1}(\mathfrak{S}_{B})]\,$ for any Borel subset $\mathfrak{S}_{B}$
of $\mathfrak{S}(\mathcal{H}_{B})$, where $\Phi ^{-1}(\mathfrak{S}_{B})$ is the pre-image of $\mathfrak{S}_{B}$ under
the map $\Phi$ \cite{H-Sh-2}. If $\mu =\{p_{i},\rho _{i}\}$ then $\Phi (\mu )=\{p_{i},\Phi
(\rho _{i})\}$.

For a given channel $\,\Phi :A\rightarrow B\,$ and an  ensemble $\mu $ in $\mathcal{P}(\mathcal{H}_{A})$
the monotonicity of the quantum relative entropy implies that
\begin{equation*}
\chi (\Phi (\mu ))\leq \chi (\mu ).
\end{equation*}%
Thus, the decrease of the Holevo quantity
\begin{equation*}
\mathrm{\Delta} ^{\!\Phi }\chi (\mu )\doteq \chi (\mu )-\chi (\Phi (\mu ))
\end{equation*}%
(called the entropic disturbance in \cite{ED-1,ED-2}) is a nonnegative function on the set of generalized ensembles with a finite
value of $\chi (\Phi (\mu ))$.

In \cite{HQL} it is shown that the function $\,\mu\mapsto\mathrm{\Delta} ^{\!\Phi }\chi(\mu)$ is lower semicontinuous on the set
$\,\{\hspace{1pt}\mu\in\mathcal{P}(\mathcal{H}_A)\,|\,\chi(\Phi(\mu))<+\infty\hspace{1pt}\}$ for any given channel $\Phi$.
By using Theorem \ref{main-t}B one can strengthen this results as follows.\smallskip

\begin{theorem}
\label{chi-loss-ls} \emph{The function
$(\Phi,\mu) \mapsto \mathrm{\Delta}^{\!\Phi}\chi(\mu)$ is lower
semicontinuous on the set
$$
\{\hspace{1pt}(\Phi,\mu)\in \F(A,B)\times \mathcal{P}(\mathcal{H}_A)\,|\,\chi(\Phi(\mu))<+\infty\hspace{1pt}\},
$$
where $\F(A,B)$ is the set of all quantum channels form $A$ to $B$ equipped with the strong convergence topology.}
\end{theorem}\smallskip

Theorem \ref{chi-loss-ls} states that
$$
\liminf_{n\rightarrow\infty}\mathrm{\Delta}^{\!\Phi_n}\chi(\mu_n)\geq\mathrm{\Delta}^{\!\Phi_0}\chi(\mu_0)
$$
for any sequences $\{\Phi_n\}\subset\F(A,B)$ and $\{\mu_n\}\subset\mathcal{P}(\mathcal{H}_A)$ converging, respectively, to a channel $\Phi_0$ and an ensemble $\mu_0$ provided that $\chi(\Phi_n(\mu_n))<+\infty$ for all $n\geq0$ (otherwise $\mathrm{\Delta}^{\!\Phi_n}\chi(\mu_n)$ is not defined).\smallskip

\emph{Proof.} Let $\{\Phi_n\}\subset\F(A,B)$ and $\{\mu_n\}\subset\mathcal{P}(\mathcal{H}_A)$ be sequences
converging, respectively, to a channel $\Phi_0$ and an ensemble $\mu_0$ such that $\chi(\Phi_n(\mu_n))<+\infty$ for all $n\geq0$.
By Theorem \ref{main-t}B there exist a separable Hilbert space $\H_E$ and a sequence of isometries $V_n:\H_A\rightarrow\H_{BE}$ strongly converging to an isometry $V_0:\H_A\rightarrow\H_{BE}$ such that $\,\Phi_n(\rho)=\Tr_E V_n\rho V^*_n\,$ for all $\,n\geq0$. Let $\nu_n$ be the image of the ensemble $\mu_n$ under the isometric channel $\rho\mapsto V_n\rho V^*_n$ for all $\,n\geq0$. It is easy to see that $\nu_n\in\mathcal{P}(\mathcal{H}_{BE})$ and that $\chi(\nu_n)=\chi(\mu_n)$ for all $\,n\geq0$. By using the arguments from the proof of Lemma 1 in \cite{AQC} one can show the weak convergence of the sequence $\{\nu_n\}$ to the ensemble $\nu_0$. Denote by $\mathrm{\Theta}$ the channel $\Tr_E(\cdot)$ from $BE$ to $B$. Then $\Phi_n(\mu_n)=\mathrm{\Theta}(\nu_n)$ and hence
$\mathrm{\Delta}^{\!\Phi_n}\chi(\mu_n)=\mathrm{\Delta}^{\!\mathrm{\Theta}}\chi(\nu_n)$ for all $\,n\geq0$. So, by Theorem 1 in \cite{HQL} (applied to the channel $\mathrm{\Theta}$) we have
$$
\liminf_{n\rightarrow\infty}\mathrm{\Delta}^{\!\Phi_n}\chi(\mu_n)=\liminf_{n\rightarrow\infty}\mathrm{\Delta}^{\!\mathrm{\Theta}}\chi(\nu_n)\geq
\mathrm{\Delta}^{\!\mathrm{\Theta}}\chi(\nu_0)=\mathrm{\Delta}^{\!\Phi_0}\chi(\mu_0).\;\; \square
$$

\begin{remark}\label{chi-loss-ls-r} Theorem \ref{chi-loss-ls} implies, in particular, that for any input ensemble $\mu$
with finite $\chi(\mu)$ and any $C\geq0$ the set of all quantum channels $\Phi$ such that
$$
\chi (\Phi (\mu ))\geq \chi (\mu )-C
$$
is closed w.r.t. the strong convergence. This result can be used
to show the closedness of the set of all quantum channels reversible
with respect to a given family of input states.
\end{remark}

A quantum channel $\Phi:A\rightarrow B$ is called \emph{reversible}
with respect to a family $\S$ of states in $\S(\H_A)$ if there exists a quantum
channel $\,\Psi:B\rightarrow A$ such that
$\,\rho=\Psi\circ\Phi(\rho)\,$ for all $\,\rho\in\S$  \cite{J-rev,O&Co}.\footnote{This
property is also called sufficiency of the channel $\Phi$ for the family $\S$ \cite{P-sqc,J&P}.}\smallskip

\begin{corollary}\label{chi-loss-ls-c} \emph{The set of all quantum channels between  quantum systems $A$ and $B$
reversible w.r.t. a given family $\S$ of states in $\S(\H_A)$ is closed  w.r.t. the strong convergence topology.}
\end{corollary}\smallskip

\emph{Proof.} Since the reversibility of a channel $\Phi$ w.r.t. an uncountable family $\S$ is equivalent to the
reversibility of this  channel $\Phi$ w.r.t. any dense countable subfamily of $\S$, we may assume that
$\S=\{\rho_i\}$ is a countable set of states. By the Petz theorem (cf. \cite{P-sqc,J&P}) a channel $\Phi$ is reversible
with respect to a family $\S=\{\rho_i\}$ if and only if
$$
\chi (\{p_{i},\Phi(\rho _{i})\})=\chi (\{p_{i},\rho _{i}\})
$$
for any probability distribution $\{p_{i}\}$ such that $\chi (\{p_{i},\rho _{i}\})<+\infty$
(in particular, for any probability distribution $\{p_{i}\}$ with finite Shannon entropy).

Thus, the assertion of the corollary  follows from Remark \ref{chi-loss-ls-r} with $C=0$. $\square$

\subsection{On continuity of the output Holevo quantity of a channel and of a complementary channel}

The Holevo quantity $\chi(\Phi(\mu))$ of the image of an input ensemble $\mu$ under a quantum channel $\Phi$ (in what follows we will call it the output Holevo quantity) plays a basic role in studying the classical capacity of this channel \cite{H-SCI,Wilde,H-c-w-c}. In quantitative analysis of the private classical capacity of a quantum channel it is necessary to deal with the output Holevo quantity a complementary channel, i.e.  with the quantity $\chi(\widehat{\Phi}(\mu))$ giving an upper bound on classical information obtained by eavesdropper \cite{H-SCI,Dev}. Since different realizations of a complementary channel are isometrically equivalent, the quantity $\chi(\widehat{\Phi}(\mu))$ is uniquely defined for any given channel $\Phi$ and input ensemble $\mu$ \cite{H-c-ch}.

Theorem \ref{chi-loss-ls} and Corollary \ref{main-c} imply the following continuity condition for the functions $(\Phi,\mu)\mapsto\chi(\mathrm{\Phi}(\mu))$ and $(\Phi,\mu)\mapsto\chi(\mathrm{\widehat{\Phi}}(\mu))$.\smallskip

\begin{proposition}\label{chi-loss-c}  \emph{If the function $\mu\mapsto\chi(\mu)$ is
continuous on a set $\P_0\subseteq\P(\H_A)$ then the functions $(\Phi,\mu)\mapsto\chi(\Phi(\mu))$ and
$(\Phi,\mu)\mapsto\chi(\mathrm{\widehat{\Phi}}(\mu))$ are continuous on the set $\F(A,B)\times \P_0$.}
\smallskip

\emph{This holds, in particular, if $\,\P_0=\{\shs\mu\in \P(\H_A)\,|\, \bar{\rho}(\mu)\in\S_0\}$, where $\,\S_0$ is any subset of $\S(\H_A)$
on which the von Neumann entropy $H(\rho)$ is continuous.}
\end{proposition}\smallskip

In other words,  Proposition \ref{chi-loss-c}  states that for any sequence $\{\mu_n\}$ of ensembles weakly converging an ensemble $\mu_0$
such that
\begin{equation}\label{imp-c}
\lim_{n\rightarrow\infty}\chi(\mu_n)=\chi(\mu_0)<+\infty
\end{equation}
and \emph{arbitrary} sequence $\{\Phi_n\}$ of channels strongly converging to \emph{any} channel $\Phi_0$ we have
\begin{equation*}%\label{imp}
\lim_{n\rightarrow\infty}\chi(\Phi_n(\mu_n))=\chi(\Phi_0(\mu_0))\;\;\textrm{and}
\;\;\lim_{n\rightarrow\infty}\chi(\mathrm{\widehat{\Phi}}_n(\mu_n))=\chi(\mathrm{\widehat{\Phi}}_0(\mu_0)).
\end{equation*}
The last part of
Proposition \ref{chi-loss-c} states that condition (\ref{imp-c}) can be replaced by the more easily verified
condition $\lim\limits_{n\rightarrow\infty}H(\bar{\rho}(\mu_n))=H(\bar{\rho}(\mu_0))<+\infty$.\smallskip

\emph{Proof.} By using Proposition 1 in \cite{H-Sh-2} and the arguments from the proof of Lemma 1 in \cite{AQC}
it is easy to show that $(\Phi,\mu)\mapsto \chi(\Phi(\mu))$ is a lower semicontinuous function on the set $\F(A,B)\times \P(\H_A)$.
So, by Theorem \ref{chi-loss-ls}, all the terms in the l.h.s of the equality
\begin{equation*}
\chi(\Phi(\mu))+\mathrm{\Delta}^{\!\Phi}\chi(\mu)=\chi(\mu)
\end{equation*}
are lower semicontinuous functions on the set $\F(A,B)\times \P_0$.  Hence, the continuity of $\chi(\mu)$ on a set $\P_0$ implies  continuity of
$\chi(\Phi(\mu))$ on the set $\F(A,B)\times \P_0$.

The continuity of the function $(\Phi,\mu)\mapsto\chi(\mathrm{\widehat{\Phi}}(\mu))$ on the set $\F(A,B)\times \P_0$ follows from Corollary \ref{main-c},
since the quantity $\chi(\mathrm{\widehat{\Phi}}(\mu))$ does not depend on a realization $\mathrm{\widehat{\Phi}}$ of a complementary channel.

To prove the last assertion note that
$\chi(\mu)+\int H(\rho)\mu (d\rho)=H(\bar{\rho}(\mu))$.
By Proposition 1 in \cite{H-Sh-2} and the arguments from the proof of the Theorem in  \cite{H-Sh-2} all the terms in the l.h.s
of the above equality are lower semicontinuous functions on the set $\P(\H_A)$.
Hence, the continuity of $H(\rho)$ on a set $\S_0$ implies  continuity of
$\chi(\mu)$ on the set $\P_0$ in this case. $\square$\smallskip

\begin{corollary}
\label{chi-loss-c++} \emph{Let $\mu$ be an ensemble in $\P(\H_A)$ with finite
$\chi(\mu)$. Then
\begin{equation*}%\label{imp}
\lim_{n\rightarrow\infty}\chi(\Phi_n(\mu))=\chi(\Phi_0(\mu))\;\;\textrm{and}
\;\;\lim_{n\rightarrow\infty}\chi(\mathrm{\widehat{\Phi}}_n(\mu))=\chi(\mathrm{\widehat{\Phi}}_0(\mu))
\end{equation*}
for arbitrary sequence $\{\Phi_n\}$ of channels strongly converging to any channel $\Phi_0$.}
\end{corollary}\smallskip

The property stated in Corollary \ref{chi-loss-c++} can be treated as stability (robustness) of the quantities $\chi(\Phi(\mu))$ and $\chi(\mathrm{\widehat{\Phi}}(\mu))$ with respect to all physical perturbations of a channel $\Phi$. Previously, the similar property was established for other two important characteristics of a channel: the quantum mutual information $I(\Phi,\rho)$ and the coherent information $I_{\rm c}(\Phi,\rho)$ \cite[Proposition 10]{CMI}.
\medskip

I am grateful to Frederik vom Ende for the valuable communications which led to the appearance of Section 6 of this paper.  I am
also grateful to A.S.Holevo, G.G.Amosov, V.Zh.Sakbaev, T.V.Shulman and M.M.Wilde for useful discussion.

Special thanks to the  participants of the workshop "Quantum information, statistics, probability", September, 2018, Steklov Mathematical Institite, Moscow, for the useful comments and suggestions which led to essential improvements of the initial version of this paper.


\begin{thebibliography}{99}

\bibitem{St} W.F.Stinespring, "Positive functions on $C^*$-algebras", Proc. Amer. Math. Soc. \textbf{6}:2, 211-216 (1955).


\bibitem{H-SCI} A.S.Holevo, "Quantum systems, channels, information.
A mathematical introduction", Berlin, DeGruyter (2012)

\bibitem{Wilde} M.M.Wilde, "Quantum Information Theory", Cambridge, UK: Cambridge Univ. Press, (2013).

\bibitem{B&Co} V.P.Belavkin, G.M.D’Ariano, M.Raginsky,"Operational Distance and Fidelity
for Quantum Channels", J.Math.Phys. \textbf{46},  062106 (2005).


\bibitem{Kr&W+} D.Kretschmann, D.Schlingemann, R.F.Werner, "The Information-Disturbance Tradeoff and the Continuity of Stinespring's Representation",  arXiv:quant-ph/0605009 (2006).

\bibitem{Kr&W} D.Kretschmann, D.Schlingemann, R.F.Werner, "A Continuity
Theorem for Stinespring's Dilation", J. Funct. Anal. \textbf{255}(8), 1889-1904 (2008).

\bibitem{Kit} D.Aharonov, A.Kitaev, N.Nisan, "Quantum circuits with mixed states", Proc. 30th STOC,
pp. 20-30, ACM Press, 1998.

\bibitem{SCT} M.E.Shirokov, "Energy-constrained diamond norms and their use in quantum information theory", Problems of Information Transmission \textbf{54}(1), 20-33 (2018).

\bibitem{W-EBN}  A.Winter, "Energy-Constrained Diamond Norm with Applications to the Uniform Continuity of Continuous
Variable Channel Capacities", arXiv:1712.10267 (2017).

\bibitem{AQC} M.E.Shirokov, A.S.Holevo, "On approximation of
infinite dimensional quantum channels", Problems of Information
Transmission \textbf{44}:2, 3-22 (2008).

\bibitem{Wilde+} M.M.Wilde, "Strong and uniform convergence in the teleportation simulation of bosonic Gaussian channels", Physical Review A, \textbf{97}:6, 062305 (2018).

\bibitem{CID} M.E.Shirokov, "Uniform continuity bounds for information
characteristics of quantum channels depending
on input dimension and on input energy",  J.Phys. A: Math. Theor  \textbf{52}, 014001 (31pp) (2019).

\bibitem{Kraus} K.Kraus, "States, Effects and Operations: Fundamental Notions of Quantum Theory", Springer Verlag (1983).

\bibitem{HQL} M.E.Shirokov, A.S.Holevo, "On lower semicontinuity of the entropic disturbance and its applications in quantum information theory", Izv. Math. \textbf{81}:5, 1044-1060 (2017).

\bibitem{W} A.Wehrl, "General properties of entropy", Rev. Mod. Phys. \textbf{50},
221-250, (1978).

\bibitem{Paul} V.Paulsen, "Completely Bounded Maps and Operator Algebras", Cambridge University Press (2003).

\bibitem{Pir} S.Pirandola, R.Laurenza, C.Ottaviani, and L.Banchi, "Fundamental Limits of
Repeaterless Quantum Communications", Nat. Commun.,~\textbf{8} 15043 (2017).

\bibitem{Datta} S.Becker, N.Datta, "Convergence rates for quantum evolution and entropic continuity bounds in infinite dimensions", arXiv:1810.00863 (2018).

\bibitem{QDS} M.E.Shirokov, A.S.Holevo, "Energy-constrained diamond norms and quantum dynamical semigroups", Lobachevskii Journal of Mathematics,
\textbf{40}(10), 1569-1586 (2019); arXiv:1812.07447.

\bibitem{H-c-w-c} A.S.Holevo, "Entanglement Assisted Capacities of Constrained Quantum Channels", Probability Theory and Applications. \textbf{48}(2), 243-255.

\bibitem{H-c-ch} A.S.Holevo, "On complementary channels and the additivity problem", Probab. Theory and Appl.,
\textbf{51}, 133-143 (2005).

\bibitem{R&S} M.Reed, B.Simon, "Methods of Modern Mathematical Physics", Vol I. Functional Analysis.  Academic Press Inc. (1980).

\bibitem{UD+} F.Caruso, J.Eisert, V.Giovannetti, A.S.Holevo, "The optimal unitary dilation for bosonic Gaussian channels",
Phys. Rev. A \textbf{84}, 022306 (2011).

\bibitem{B&R} O.Bratteli, D.W.Robinson, "Operators algebras and quantum statistical mechanics", vol.I, Springer Verlag,
 New York-Heidelberg-Berlin (1979).

\bibitem{SSC} M.E.Shirokov "Strong-$*$  convergence of quantum channels", arXiv:1802.05632.

\bibitem{Bil} P.Billingsley, "Convergence of probability measures", John
Willey and Sons. Inc., New York-London-Sydney-Toronto  (1968).

\bibitem{Par} K.R.Parthasarathy, "Probability measures on metric spaces",
Academic Press, New York and London (1967).

\bibitem{H-Sh-2} A.S.Holevo, M.E.Shirokov "Continuous ensembles and the $\chi$-capacity of infinite dimensional channels", Probab. Theory and Appl. \textbf{50}:1, 86-98 (2005).

\bibitem{H-73} A.S.Holevo, "Bounds for the quantity of information transmitted by a quantum communication channel",
Probl. Inf. Transm. (USSR) \textbf{9}, 177-183 (1973).


\bibitem{ED-1} F.Buscemi, M.Horodecki, "Towards a
unified approach to information-disturbance tradeoffs in
quantum measurements",  Open Systems and Information
Dynamics, \textbf{16}(01), 29–48 (2009).

\bibitem{ED-2} F.Buscemi, S.Das, M.M.Wilde, "Approximate reversibility in the context of entropy gain, information gain, and complete positivity",
Physical Review A \textbf{93}:6,  062314 (2016).

\bibitem{J-rev} A.Jencova, "Reversibility conditions for quantum operations",  Rev. Math. Phys. \textbf{24},  1250016 (2012).

\bibitem{O&Co} T.Ogawa, A.Sasaki, M.Iwamoto, H.Yamamoto, "Quantum Secret Sharing Schemes and Reversibility of Quantum Operations", Phys. Rev. A \textbf{72}, 032318 (2005).

\bibitem{P-sqc} D.Petz., "Sufficiency of channels over von Neumann algebras", Quart. J. Math. Oxford
Ser. (2) \textbf{39}:153, 97-108 (1988).

\bibitem{J&P} A.Jencova, D.Petz, "Sufficiency in quantum statistical inference",
Commun. Math. Phys. \textbf{263}, 259-276, (2006).


\bibitem{Dev} I.Devetak, "The private classical capacity and quantum capacity of a quantum channel", IEEE Transaction on Information Theory
\textbf{51}:1, 44-55 (2005).

\bibitem{CMI} M.E.Shirokov, "Measures of correlations in
infinite-dimensional quantum systems",
Sbornik: Mathematics \textbf{207}:5, 724-768 (2016).

\end{thebibliography}
\end{document}